# A New Scientific Revolution at the Horizon?

Gilles Cohen-Tannoudji[*] et Sylvain Hudlet[**]

[*]Laboratoire de Recherche sur les Sciences de la Matière (LARSIM) CEA-Saclay, F91191 Gif sur Yvette Cedex
[**]Université de Technologie de Troyes, Département Matériaux, 12, rue Marie Curie, BP 2060, 10010 Troyes Cedex

**Abstract** At this beginning of the 21st century, the situation of physics is not without analogy with that which prevailed a hundred years ago, with the outset of the double scientific revolution of relativity and quanta. On the one hand, recent progress of observational cosmology makes think that one has discovered a new universal constant, perhaps as fundamental as the velocity of light or the Planck's constant, the cosmological constant, which could explain the acceleration of the expansion of the universe. On the other hand, just like the efforts of Planck and Einstein to reconcile thermodynamics and the electromagnetic theory of light led to the operational beginning of quantum physics, the unexpected discovery of bonds between thermodynamics and general relativity makes to foresee new concepts, perhaps heralding a new scientific revolution, like that of holography and leads to consider a "thermodynamic route towards quantum cosmology." We will discuss the possible implications of these observational and theoretical developments.

## 1/ Introduction

In these same places, four years ago, we celebrated the centennial of the miraculous year of Einstein during which he gave the starting point of the scientific revolution of the 20th century. Today, we celebrate the four hundredth anniversary of the use by Galileo of his telescope which, also, marks the starting point of a scientific revolution, the one which saw being born then developing modern science, since the theory of the universal gravitation by Newton until the apogee of classical physics at the end of the 19th century. Whereas the scientific revolution of the 20th century leads to an apogee comparable with that of classical physics, the progress achieved by observational cosmology and very lively theoretical developments seem to suggest that a new scientific revolution appears at the horizon. It is what we will try to explain in this conference.

At the end of the 19th century, the apogee of classical physics consisted of three theories which concretized significant syntheses or unifications and which made it possible to model in a satisfactory way the whole of the then observable phenomena: the *electromagnetic theory of light* by Faraday, Maxwell and Hertz which unified the electric, magnetic and optical phenomena, the *theory of universal gravitation* by Galileo and Newton which unified terrestrial mechanics and celestial mechanics and the conjunction of *analytical mechanics* by Lagrange and Hamilton, of the *kinetic theory* of matter and *statistical thermodynamics* by Maxwell and Boltzmann leading to the unification of rational mechanics with the atomistic conception of ancient Greek philosophers.

The overall structure of the theoretical framework of classical physics is maintained in the physics of the 20th century. This framework now includes three theories taking each one into account a couple of dimensioned universal constants which prolong the theories of the framework of classical physics: *quantum field theory* (constants $\hbar$ and $c$) which prolongs the electromagnetic theory of light and is used as a basis for the *standard model* of the physics of particles and their non-gravitational fundamental interactions, the *general theory of relativity* (constants $G$ and $c$) which prolongs the theory of universal gravitation by Newton and is used as a basis for the *standard cosmological model*, and *quantum statistics* (constants $\hbar$ and $k$) which prolongs analytical mechanics and statistical thermodynamics and which is used as a basis for the phenomenological consolidation of the standard models of particle physics and cosmology.

Which elements of comparison do exist between the pre-revolutionary situation of the end of the 19th century and today? Lord Kelvin (William Thomson), analyzing in 1900 the field of investigation of physics, announced that it was about to be completed except for two "small clouds", of which he thought that they would require only some adjustments to be reabsorbed. It was about the failure of the detection of the earthmoving in ether (experiment of Michelson and Morley) and of the absence of theoretical explanation to the observed black body spectrum. The resolution of the first difficulty thus gave place to the special theory of relativity and then to the general theory of relativity and that of the second one led to quantum physics.

Nowadays, it is regularly written that contemporary physics, despite of several attempts still fails establishing a theory which would contain at the same time general relativity and quantum physics. Is it in addition possible to reinstate in physics the recent projections of observational cosmology? Will that make it possible to reach a unified comprehension of physics? We will see that there seem today to emerge some completely new approaches of these questions, some having been proposed very recently. If they have as a common point, with the resolutions of the questions raised by Lord Kelvin, to call into question the idea of space and to include statistical thermodynamics, they especially have the advantage of showing in what these two questions are dependent.

## 2/ Quantum field theory and the standard model of particle physics

### 2.1 Quantum field theory

Quantum field theory (QFT) carries out the marriage of special relativity (taking into account $c$) and quantum mechanics (taking into account $\hbar$). Due to the weakness of the gravitational interaction at experimentally accessible energies, it postpones the marriage of quantum mechanics and general relativity.

The constants $c$ and $\hbar$ translate fundamental limitation principles which are obeyed by QFT. The vacuum speed of light $c$ is the universal constant translating, always and everywhere, the impossibility of instantaneous action at a distance. The existence of an elementary quantum of action $\hbar$ excludes any subdivision of individual quantum processes, which must be neither treated, individually, as predictable or reproducible events. For these two principles to be obeyed, the quantum fields are:
- Relativistic fields, i.e. defined at each point of the Minkowski space-time;
- Quantum fields, i.e. fields of operators acting in a Hilbert space and causing events of emission or absorption of energy quanta;
- Energy quanta are particles or *antiparticles* (a feature that solves the problem of negative energies).;
- Interactions are described by means of local couplings, i.e. products of fields evaluated at the same point of space-time;
- The *vacuum* is the fundamental or ground state of the system of quantum fields in which the number of energy quanta is null, but where the quantum fields are affected by quantum fluctuations (for example the formation followed by the annihilation of a particle-antiparticle pair).

The locality of the interaction couplings induces a singularity in QFT: the integrals from which one obtains the transition amplitudes are divergent. It is the theory of *renormalization* which makes it possible to overcome this difficulty: a theory is known as *renormalisable* if all observable physical quantities can be expressed without infinity in terms of parameters that depend on energy, that are redefined (it is said "renormalized") by the interaction. A renormalisable theory is *predictive*: the parameters on which it depends are in a finite number and they can be determined experimentally, but it is not "fundamental", since the value of the parameters depends on the resolution (i.e. on the available energy). This theory is *effective* because the dependence in energy of the parameters is predictable, thanks to the of the *renormalization group equations* (RGE).

## 2.2 The standard model of particle physics

The standard model of particle physics is the result of the application of QFT to the non-gravitational fundamental interactions, namely the electromagnetic interaction and the strong and weak nuclear interactions.

The quantum and relativistic theory of the electromagnetic interaction, known as Quantum Electro-Dynamics (QED) was, at the end of the Forties, the first stage of the construction of the standard model. It is a theory in which the effects of the quantum corrections are calculable (because the theory is renormalisable) and measurable in atomic physics at low energy. The agreement between the theoretical predictions and the experimental data exceeded all the hopes, so that this theory was used as a model for the development of the theory of the other fundamental interactions.

This development was made possible, on the one hand, by the discovery in the Sixties, of a new level of elementarity, the level of *quarks*, the elementary constituents of the hadrons, the particles which take part in all fundamental interactions including the strong interaction, and on the other hand by the identification of the property of symmetry that is essential in QED and likely to be generalized to other interactions, *gauge invariance*. The standard model then includes:

- Quantum Chromo-Dynamics (QCD), a renormalisable theory with gauge invariance describing the high energy interactions of quarks by exchange of *gluons*;
- The *electroweak unified theory*, a renormalisable gauge invariant theory ideally describing the interactions of the quarks and the *leptons* (i. e. the fermions which do not take part in the strong interaction), if they were massless by the exchange of the photon and of the *intermediate vector bosons* of the weak interaction, if they were massless;
- The *Higgs mechanism* which induces the spontaneous breaking of the electroweak symmetry and makes massive the quarks, the charged leptons and the intermediate vector bosons while preserving the renormalisability of the electroweak unified theory;
- This mechanism implies the existence of at least a not yet discovered particle, the *Higgs boson*, the research of which is the top priority assigned to the Large Hadron Collider (LHC), the collider that was commissioned at CERN at the end of 2009. Except for this last missing link, the overall agreement between the theoretical predictions of the standard model and the whole of the experimental data up to energies about the hundred of GeV is satisfactory (of the order of the percent).

## 2.3 Which new physics beyond the standard model?

The standard model which consists of effective theories is not insuperable. The parameters of which it depends, as those which measure the intensity of the interactions at the elementary level, are not constant, they depend on energy. The standard model which works well at the highest currently accessible energies can very well be embedded, at higher energy, in a theory which would include it and would give it again as a low energy approximation. As it was noted that the intensities of the non gravitational interactions seem to converge at an energy of about $10^{15}$ GeV, it is tempting to suppose that at this energy these interactions could be unified in a grand unification theory (GUT). In such a theory the proton would be unstable and certain problems unsolved within the standard model, like the breaking of the matter-antimatter symmetry, or neutrino masses could find a solution. But such a theory would suppose the existence of a new Higgs mechanism occurring at an energy $10^{13}$ times higher than the energy of the electroweak symmetry breaking. The articulation of two Higgs mechanisms at such different energies poses a very difficult problem, of which a solution could be found thanks to a new property of symmetry, *supersymmetry*. One has thus developed an extension of the standard model, the minimal supersymmetric standard model (MSSM) with which one would recover all the assets of the standard model at energies lower than that of the LHC, but which would lead to the prediction of new observable effects at the LHC energies, such as the existence of many new particles. The way of GUT is considered by particle physicists as the direct way towards the reconciliation of quantum physics and general relativity, within a quantum theory of gravitation, whose field of validity would be that of the scales of Planck

$$\text{length } l_P = \left(\frac{hG}{c^3}\right)^{1/2} \approx 10^{-35} m$$

$$\text{time } t_P = \left(\frac{hG}{c^5}\right)^{1/2} \approx 10^{-43} s$$

$$\text{mass } m_P = \left(\frac{hc}{G}\right)^{1/2} \approx 10^{19} GeV/c^2$$

But, as we are going to show now, another way seems to open in the direction of a quantum theory of gravitation, that of thermodynamics.

## 3/ General relativity and the standard cosmological model

### 3.1 General covariance, principle of equivalence and a geometrical theory of gravitation

The theory of relativity was developed by Einstein in two stages: in 1905, special relativity integrates the Galilean principle of relativity (equivalence of inertial reference frames moving in relative uniform rectilinear motion) and in 1916, general relativity extends the principle of relativity to arbitrary changes of reference frames[1].

To generalize the principle of relativity to arbitrary changes of reference frames (principle of general covariance), Einstein makes a detour through the theory of universal gravitation: starting from the independence of the acceleration communicated to a body by gravitation with respect to the mass and to the other properties of this body, independence which had been noticed by Newton and which he promotes to the status of the *principle of equivalence*, Einstein shows that:
(i) An arbitrary change of reference frame can be replaced, *locally* (i.e. in an infinitesimal domain of space-time), by an adequate gravitational field and
(ii) The gravitational field can be replaced, locally, by an adequate change of reference frame.

He thus arrives at a geometrical theory of gravitation whose fundamental equation connects the Ricci-Einstein tensor $G_{\mu\nu}$ related to the non-Euclidean geometry of space-time to the energy-momentum tensor $T_{\mu\nu}$ describing in a phenomenological way the properties of matter. The proportionality constant relating these two tensors is fixed in such a way that one recovers Newtonian gravitation theory at the non-relativistic limit: $G_{\mu\nu} = 8\pi G/c^4 \ T_{\mu\nu}$. General covariance implies that only events of space-time coincidence are observable. Arrived at his equation, Einstein made the following comment about it: "the theory avoids all the defects which we reproached to the foundation of classical mechanics. It is sufficient, as far as we know, for the representation of the observed facts in celestial mechanics. But it resembles a building whose one wing is built of fine marble (first member of the equation) and the other of wood of lower quality (second member of the equation). The phenomenological representation of the matter compensates, actually, only very imperfectly a representation which would correspond to all the known properties of matter[2]."

General relativity implies departure from the theory of Newton only when the gravitational fields are strong. The importance of such effects can be evaluated simply by determining the escape velocity, i.e. the speed above which a test body can escape from the gravitational field generated by a planet or a star. If this speed is not negligible as compared to the velocity of light, then the Euclidean geometry, or more precisely the Minkowskian geometry of space-time of special relativity, is not relevant any more in the description of the laws of mechanics. The curvature of space-time must be taken into account. It remains to appreciate the importance of gravity. In the case of a spherical object of radius *R*, the escape velocity is given by $2GM/R$. On Earth, it is about 11 km/s, so the relativistic effects are not very visible (although it is already necessary to introduce corrections due to general relativity into the very precise synchronization of the GPS). General relativity effects become significant only on stellar scales.

---

[1] Albert Einstein, "Die Grundlage DER allgemeinen Relativitätstheorie", Annalen der Physik. Vol. XLIX, 1916, p. 769-822

[2] A. Einstein, Physics and Reality, Franklin Institute Journal, vol.221, n° 3, March 1936

In the article of synthesis appeared in 1916, in which general relativity is completely formalized, Einstein indicates three possible validations of his theory: the advance of the perihelion of the elliptic orbit of Mercury around the sun (of 43" per century!), a shift of the spectral lines emitted by massive stars and the deviation of the rays in the vicinity of the sun (which was measured a little time later, at the time of the eclipse in 1919, as envisaged by relativity).

### 3.2 General relativity and cosmology

If it thus appears that significant masses are necessary to lead to departures from the Newtonian theory of gravity, an additional argument makes it possible to understand the reason why such effects can show up at cosmological scales. Just like in classical gravitation, general relativity treats only positive masses (or in an equivalent way their energy content $E = mc^2$). As a consequence the effects of curvature are additive. Thus, with an average density of matter in the universe, the theory can associate an average curvature of space-time. General relativity thus appears as being naturally connected to cosmology, the science which has as an aim the description of the whole universe. A universal description of the world is then possible because the universe appears homogeneous and isotropic at long distance. This formalization thus rests on the cosmological principle that stipulates that there exists a universal time, but that there exists neither a geometrical position nor a direction that are privileged in space.

Inaugurating this way of thinking, Einstein realizes as soon as 1917 that the universe of general relativity is unstable. From the additive character of the gravitational attraction that we have just mentioned, follows the fact that uniformly distributed matter at rest cannot be in a stable equilibrium: spontaneously, the Universe should collapse to a point. To mitigate that, and to thus ensure cosmos a total immutability, Einstein adds to the equation of general relativity, an extra term, perfectly respecting its general covariance, which he denotes $\Lambda$, and which has been thereafter called the *cosmological constant*. The positive value of this constant physically translates the effect of a repulsive internal pressure able to counterbalance the attractive action of gravitation. Within the framework of the cosmological principle, the solution of the Einstein equation thus obtained[3] corresponds to a static and elliptic universe (finite, but without edge just as the two dimensional surface of a sphere), in which a light ray would return to its departure point. The first component of this equation makes it possible to connect the radius of the universe *R*, the cosmological constant $\Lambda$ and the density of matter $\rho_0$

$$\underbrace{\overbrace{\frac{c^2}{R^2}}^{\text{Space curvature}} - \frac{\Lambda c^2}{3}}_{\text{metric}} = \underbrace{\frac{8\pi G}{3} \rho_0}_{\text{Energy}}$$

Einstein is all the more satisfied with the addition of this constant that it leads to a cosmological model satisfying what he calls the Mach's principle which requires that the gravitational field as well as inertia are completely determined by the energy content of the universe. In such a model the Mach's principle is indeed respected because the cosmological constant, that is purely geometrical (its dimensional content is that of the inverse of the square of a length) is independent of the energy content of the universe, a content that is completely described by the density $\rho_0$, since the universe is finite.

---

[3] Albert Einstein, « Kosmologische Betrachtungen zur allgemeinen Relativitaetstheorie », Sitzungsberichte der Preussischen Akademie der Wissenschaften zu Berlin, 1917, p 142-152
French translation in « Albert Einstein, Œuvres choisies », tome 3, p 88, Ed. du Seuil

It remains that this cosmological constant revealed, as shown by de Sitter a few months later[4], a formal solution to the equations of relativity with no right hand side, completely unacceptable with respect to the Mach's principle: a universe of elliptic geometry void of matter but not deprived of gravitational field! In fact such a universe would be stable because it would not contain any matter. If particles, of mass sufficiently low not to affect the global geometry, are deposited in a given region of space, the effect of the cosmological constant will be to move them away towards an *event horizon* (located at a finite distance). The presence of this horizon was analyzed by Einstein as a singularity without physical counterpart.

The interpretation of the role of the cosmological term can in fact be reduced to its position in the equation of Einstein. If it is in the left hand side (related to the metric tensor), the analysis of de Sitter indicates that the Mach's principle is not respected, whereas if it is in the right hand side (related to the energy-momentum tensor), its origin is to be brought back to the microscopic interactions but then the effect of the repulsion remains to be explained.

During the Twenties, the problem of the cosmological constant appeared more and more severe. It was thus shown, on the one hand that even the idea of Einstein of relying on a cosmological constant leads to a universe that is unstable, and on the other hand, that the singularity of the horizon of the de Sitter model which Einstein denounced was actually not a singularity. It is with the works by Friedmann in 1922 and Lemaître in 1927 that the events took another turn[5]. Each one of these two authors showed, independently of the other, that there were also dynamical solutions to the equation of Einstein. Space-time thus has the property of being able to dilate or to contract. Mathematically, these solutions then connect the average density of matter not to the radius of the universe but rather to its scale factor $a(t)$ and its time derivative $\dot{a}(t)$. One thus has

$$\underbrace{\frac{c^2}{a(t)^2}+\frac{\dot{a}(t)^2}{a(t)^2}}_{\text{space-time curvature}}=\underbrace{\frac{8\pi G}{3}\rho_0}_{\text{energy}}$$

Such an innovation makes useless the addition of the cosmological constant.

Initially, Einstein was not very much interested in these dynamical solutions because they corresponded to open universes, for which the inertial interpretation of the Mach principle becomes problematic at infinity. He started to change his opinion at the end of the decade when observational evidence of a temporal evolution of the geometry started to come up at the horizon. This is how the "big-bang" theory (according to the name that the astronomer Fred Hoyle gave it around the middle of the 20[th] century), adding to relativity the expansion of space, produced the first valid cosmological model for which the addition of the cosmological constant was not necessary any more.

## 3.3 The standard cosmological model: from the simple big-bang model to the cosmology of concordance

It is to Edwin Hubble that one owes the first observational indices of an expansion of the universe[6]. This astronomer used for that the velocity measurement of far away distancing galaxies

---

[4] W. de Sitter, « On the Relativity of Inertia. Remarks Concerning Einstein's Latest Hypothesis », Proc.K.Akadem.Amsterdam, XIX, 1917, p 1217-1225, ; « On the Curvature of Space », Proc.K.Akadem.Amsterdam, XX, 1917, p 229-242 ; « On Einstein's Theory of Gravitation and Its Astronomical Consequences. Third Paper », Roy. Soc. Astron. Soc. Monthly Notices, LXXVIII, 1917, p 3-28

[5] A. Friedmann, « Über die Krümmung des Raumes », Z. Phys., 10 (1), 1922, p. 377-386
Abbé G. Lemaître, « Un univers de masse constante et de rayon croissant, rendant compte de la vitesse radiale des nébuleuses extra-galactiques », Ann. Soc. Sci. Brux., XLVII, série A, 1927, p 49-59

[6] E. P Hubble, « A relation between distance and radial velocity among extra-galactic nebulae », Proceedings of the National Academy of Sciences, 15, 1929, 168–173

whose position was also known (velocity is in fact given by the Doppler shift on the spectral lines, whereas the distance is known thanks to Cepheid stars, present in these galaxies and whose characteristics of absolute luminosity was known). The result which he announced in 1929 was that the speed of distancing was proportional to the distance, that is to say $v = H_0\ d$ where $H_0$ is the constant which now bears his name and which makes it possible to determine the growth rate of space in the models of Friedmann and Lemaître (the numerical determination of this constant is frightening of difficulty; today, it seems to be equal[7] to 70,5 ± 1,3 km/s/Mpc).

The very existence of a linear relation speed/distance is a proof of the expansion – and of a uniform expansion – of the universe. Its origin is easily understood by means of the following analogy made with a one dimensional model: if one draws one end of a rubber band, the other end being maintained fixed, the displacement from the fixed origin of a particular point of the rubber band will be the larger the more it is initially distant from it. The fact that the proportionality constant is a scalar adds to the property of homogeneity of space, the one of its isotropy.

To the Hubble's proof two other major evidences of the expansion of the universe are added today. The first one relates to relative abundances of the elements resulting from primordial nucleosynthesis. It has thus been noticed that the relative proportions of helium, deuterium and lithium are appreciably uniform in the whole universe. This indicates a common origin of the light elements. As those can be formed starting from protons and of neutrons only for temperatures of about $10^9$ K, one can conclude from it that it is necessary that the universe was, at a former period, hotter and thus denser. That corresponds precisely to the idea of the expansion. It is in this manner that George Gamow, Ralph Alfer and Hans Bethe proceeded to show in 1948 that the standard theory of the big-bang led to an exact forecast of relative abundances of the light elements[8].

The second proof of the big-bang model resides in the observation of the *cosmological microwave background* (CMB). This electromagnetic radiation, detected in 1965 by two radio astronomers, Arno Penzias and Robert Wilson[9], was a prediction of the big-bang theory made firstly by George Gamow, and then by Ralph Alpher and Robert Herman[10] in 1948. This radiation indeed originates in the radiative transitions from the first neutral atoms which could have been formed when the temperature of the universe became gradually rather low (about 3 000 K). Between this time (some 380 000 years after the big-bang) and today (approximately 13,7 billion years after the big-bang), the wavelength of the emitted photons then increased with the expansion of the universe so that this radiation initially lying in the visible and the ultraviolet domain is detected today in the domain of the radio waves. Perhaps this shift in wavelength constitutes the most remarkable proof of the expansion of space-time.

Let us have a pause to specify the interest for cosmology of the observation of this cosmological background. At the time of its emission, matter is not organized: atoms are in thermal equilibrium with the radiation. The maximum of intensity of the radiation can then be connected, thanks to the black body theory discovered by Planck, to the temperature of matter. It was the object of the sending of the COBE and WMAP satellites (respectively launched in 1989 and 2001) to measure with a high accuracy the space variations of the temperature of the cosmological background. This temperature is on average equal[11] to 2,726 K and its fluctuations (once are subtracted the Doppler effect caused by the earthmoving and the effects of known sources in the Milky Way) relate to the fifth figure of the preceding value. If the study of this radiation thus states that the universe was remarkably homogeneous at its beginnings, these very small fluctuations also provide significant information because the appearance of the large scale structures which one currently observes in the distribution of galaxies in the universe can be partly connected to the temperature fluctuations in the primordial matter.

---

[7] « Five-Year Wilkinson microwave anisotropy probe observations : cosmological interpretation », The Astrophysical Journal Supplement Series, 180 :330-376, 2009
[8] R. A. Alfer, H. Bethe, G. Gamow, « The Origin of Chemical Elements », Physical Review Letters, 73, 1948, p. 803-804
[9] A.A.Penzias et R.W.Wilson, « Measurement of Excess Antenna Temperature at 4080 Mc/s », Astrophysical Journal, Vol. 142, 1965, p 419-421
[10] G. Gamow, « The evolution of the universe », Nature, 162, 1948, p. 680 ; R. A. Alpher, R. Herman, « Evolution of the universe », Nature 162, 1948, p. 774
[11] « Measurement of the cosmic microwave background spectrum by the Cobe Firas », Astrophysical Journal, 420, 439 (1994)

The models of Friedmann and Lemaître make it possible to correctly describe a universe compatible with the three major evidences of the expansion which have just been given. These cosmological models however do not allow explaining that the cosmological background is so homogeneous and that the curvature of space (deduced for example from the density of matter and the growth rate provided by the measurements of the WMAP satellite) is so low. The observed homogeneity of the CMB is particularly paradoxical because it implies that two regions that have never been in the past in causal contact should have now almost exactly the same temperature. Which could thus be the reason which makes them resemble each other so much? It was proposed to explain this fact to supplement the first cosmological model which has just been presented, with a new hypothesis, the one of an initial exponential expansion called *inflation* of space-time.

This idea was initially proposed, at the beginning of the Eighties, by Alan Guth[12]. It can be exposed in the following way. In the very remote past of the universe, i.e. at temperatures ranging between $10^{27}$ K and $10^{32}$ K corresponding to times after the big bang ranging between the Planck's time ($10^{-43}$ s) and the time of the grand unification symmetry breaking ($10^{-35}$ s), a phase transition would have occurred whose causes remain to be explained. It could have been translated, in the equation of Einstein, by the appearance of a term similar to the cosmological constant. During this period, the universe would have extended in space according to an exponential of the time, so that the scale factor of the universe would have increased at least by 26 orders of magnitude. This brutal dilation would have leveled the universe while making its spatial curvature negligible. Moreover, the various areas of the sky today observed in the cosmological background would then have been causally connected in their very remote past, a fact that cannot be accounted for by the sole Hubble constant. This scenario of inflation, intellectually attractive, missed, until very recently, any observational support.

It turns out that, precisely, observational cosmology made, very recently, considerable progress in two fields: the one of the measurement of distances by means of the observation of supernovas of the type IA in remote galaxies, which made it possible to improve the determination of the temporal dependence of the scale factor of the Universe, and the one of the measurement of the cosmological background, which allowed, starting from a very detailed study of its fluctuations, to improve the determination of the various components of the energy density. Using phenomenological models, the interpretation of the data coming from these two fields converged towards what one now calls the *cosmology of concordance*, or the $\Lambda$CDM[13] model, for Lambda-Cold-Dark-Matter, which can be summarized in the following way:

– The spectrum of the fluctuations of the CMB is compatible with the inflation scenario, which thus leaves the realm of pure speculation;
– The age of the Universe is estimated at 13,7 billion years, up to a few percents;
– The date of emission of the CMB is 379 000 years after the big-bang;
– The energy density of the Universe is compatible[14] with the density, known as the *critical density*, corresponding to a spatially flat Universe: $\rho_c = \frac{3H_0^2}{8\pi G}$, where $H_0$, the Hubble constant is the inverse of the radius of the current horizon;
– The density of baryonic[15] matter lies between 4 and 6 % of the total density;
– Observations concerning the dynamics of large scale structures suggest the presence, to a total amount of 20 to 35 % of the total density, of a kind of matter called *dark matter*, which appears only by its gravitational effects;
– To complete the total energy density of the Universe, one must include another component, called *dark energy*, which represents approximately 70 % of it and which seems to have the same effect (a negative pressure accelerating the expansion) as the cosmological constant

---

[12] Guth, Alan, *The Inflationary Universe: The Quest for a New Theory of Cosmic Origins*. 1997. ISBN 0-201-32840-2 or ISBN 0-224-04448-6
[13] O. Lahav and A.R. Liddle *The cosmological parameters,* arXiv:1002.3488
[14] Its value is between 98% and 108% of the critical density
[15] Namely matter made of atoms the nuclei of which consist on protons and neutrons

introduced and then given up by Einstein. This last observation, which one can describe as a true discovery, is a complete surprise. If confirmed, it would lead to a sign heralding a new scientific revolution since it would lead to a surprising prediction about the far future of the universe: in fact with a non vanishing cosmological constant, when time goes to infinity the horizon radius goes to the length associated with the cosmological constant, the energy density goes to the dark energy density $\rho_{DE}$, the observable universe (i.e. the universe inside the horizon) becomes empty since all galaxies are beyond the horizon[16]

– $\quad t \to \infty \Rightarrow H \to H_\infty = 1/L_\Lambda ; \rho_c \to \dfrac{3}{8\pi L_\Lambda^2 G} = \rho_{DE}$

–

## 4/ Thermodynamics of black holes

### 4.1 Entropy and temperature of black holes

Already in the Seventies, some works of Penrose, Hawking, Bekenstein and Carter on black holes led to an understanding of their physics which required elements of thermodynamics. Let us come back, to explain that, to the classical expression which connects the escape velocity to the size of a massive sphere. It turns out that, since no object can exceed the velocity of light, a body of mass *M* will retain with it everything lying inside a region of radius *R* smaller than $2GM/c^2$ (the *Schwarzschild radius*). The boundary of this region is the *event horizon* of the black hole (an external observer will not see anything of what occurs inside this zone). It is the origin of the bond with thermodynamics. Contrary to classical mechanics where the motion of particles is reversible, the physics of the black hole imposes an orientation of time. This characteristic is at the very foundation of thermodynamics. Thus, as the area of the horizon of a black hole can only increase when it accretes matter, it is possible to make it play the role of an entropic variable.

Bekenstein[17] determines in 1973 the precise expression of the entropy of a black hole using arguments based on quantum physics. Let us sketch his reasoning. To increase the entropy of the black hole by the smallest amount, a bit of information *k*Ln2 (*k* is the Boltzmann's constant) is dropped into the black hole in the form of a photon of the smallest possible energy, i.e. with a the wavelength equal to the Schwarzschild radius. The corresponding increase in energy is thus given by the relation of Einstein, $\Delta E = h\nu = hc/R$, i.e. a mass increase $\Delta M = \Delta E/c^2$, corresponding to an increase in the radius of the horizon $\Delta R = 2\Delta MG/c^2 = 2hG/c^3 R$, and thus to an increase of the surface area equal to $\Delta A = 8\pi R \Delta R = 4\hbar G/c^3 = 4 A_P$, namely four times the Planck's surface area. This area increase is independent of any particular characteristics of the black hole (such as its mass). Starting from this differential evaluation, one can go up to the total entropy of the black hole by noting that the entropy of a black hole of null size being null, there is no constant of integration. One thus expects that the total entropy of the black hole is equal to $S = \eta k \operatorname{Ln} 2 A / 4 A_P$ where the unknown constant factor $\eta$ is of order 1. In addition, Hawking[18] has shown that it is also possible to associate a temperature $T = \dfrac{1}{2\pi} \dfrac{\hbar c^3}{k} \dfrac{1}{GM}$ with the black hole that quantum physics authorizes to evaporate. By writing the second principle of thermodynamics with this temperature, one can fix the value of the constant unknown factor $\eta$. One then finds $S = k \dfrac{A}{4 A_P}$ .

---

**4.2 Bekenstein's bound and holographic principle**

As the formation of a black hole is the most effective way to compress matter in a certain volume, the Bekenstein's entropy appears as the upper limit of the information which can be contained in a sphere of space time. This upper limit is expressed using the *holographic principle*[19]: "How many degrees of freedom are there in nature, at the most fundamental level? The holographic principle answers this question in terms of the area of surfaces in space-time: (…) A region with boundary of area A is fully described by no more than A/4 degrees of freedom, or about 1 bit of information per Planck area[20]."

**4.3 The problem of information non-conservation and its solution by Susskind**

As the evaporation of a black hole seems to be a non-unitary purely thermal process, Hawking[21] raises the problem of the non-conservation of information in the dynamics of black holes, which could announce an irreducible incompatibility between gravitation and quantum physics for which the principle of conservation of information (also called principle of *unitarity* of the *S*-matrix) is fundamental. L. Susskind describes in his popular science book[22] the long debate which opposed him to Hawking on this subject and the way in which he managed to solve the problem. He shows[23] that the paradox raised by Hawking is due only to the approximation, known as semi-classical, which he used in the modeling of the evaporation of the black hole and that an entirely quantum treatment of gravitation should allow to solve the paradox. Although a quantum theory of gravitation is not yet available, he proposes a model which, thanks to the holographic principle, respects the principle of conservation of information in the complete process which goes from the formation to the evaporation of a black hole: the quantum dynamics of the black hole is described by means of a unitary *S* matrix defined on the horizon of the black hole. The unitarity of this "holographic" *S* matrix is ensured by a combination of the principle of equivalence of general relativity and the principle of complementarity of quantum physics. The principle of equivalence tells us that an observer maintained outside the black hole perceives its horizon like a thermal system (namely a body black), whereas an observer in free fall in the black hole does not perceive the horizon which is a border of no return. How comes that the information perceived by the observer in free fall is not irremediably lost for the external observer? The answer suggested by Susskind to this question lies in the principle of complementarity of quantum physics: according to the above mentioned holographic principle, all information concerning the evolution of the black hole is encoded on the horizon, and in quantum physics, information residing on both sides of the horizon, which can be accessed only under contradictory conditions of detection (that of the observer in free fall and that of the external observer), is complementary, like are, for example, the dynamic variables constrained by the inequalities of Heisenberg.

---

[19] The reasoning based on the arguments of Bekenstein and Hawking which we presented here is primarily heuristic. It is within the framework of the superstring theory that was carried out in a rigorous way the calculation of the entropy associated with the horizon of a black hole, see E Witten, *Anti-in Sitter space and holography*, Adv. Theor. Math Phys. 2, 253. (ArXiv:hep-th/9802150)

[20] R. Bousso, *The holographic principle,* Rev.Mod.Phys.74:825-874 (2002) (arXiv:hep-th/0203101)

[21] S. Hawking, *Breakdown of predictability in gravitational collapse,* Phys. Rev. D 14, 2460 (1976)

[22] L. Susskind, *The Black Hole War – My Battle with Stephen Hawking to Make the World Safe for Quantum Mechanics.* Little, Brown and Compay, New York, Boston and London, 2008. .

[23] L. Susskind, L. Thorlacius and J. Uglum, *The stretched horizon and black hole complementarily,* Phys. Rev. D 48, 3743 (1993) (arXiv:hep-th/9306069). Indépendamment de Susskind et presque simultanément G. 't Hooft a abouti à des conclusions similaires : voir  G. 't Hooft, *Dimensional reduction in quantum gravity* arXiv:gr-qc/9310026

## 5/ Thermodynamics and gravity: gravity as an emergent phenomenon

### 5.1 A thermodynamic route towards quantum cosmology

The idea that gravity can be described as an emergent phenomenon has a long history which originates with the work of Sakharov[24]. Gravity-thermodynamics connection was discovered by Jacobson[25] who used the proportionality of the entropy to the area of the horizon and a classical thermodynamical identity to assimilate the Einstein's equation to an equation of state. The implications of this connection were thoroughly analyzed by Padmanabhan[26]. In reference[27] he presents the guiding principles of his program and the stages of what would be a thermodynamic route towards quantum cosmology:

1. The horizons are inevitable in the theory and they always depend on the observer;
2. The thermal nature of the horizons cannot occur without space-time having a microstructure
3. All observers have the right to describe physics using an effective theory based on the variables to which they have accesses;
4. The problem of the cosmological constant (why is it so small?) is due only to our bad understanding of the nature of gravitation. This problem cannot be solved in a theory arising from an action which (i) is generally covariant, (ii) uses as dynamic variables the components of metric and (iii) comprises a mater sector whose energy is defined up to an additive constant;
5. Gravity is an emergent phenomenon, which means that the components of the metric tensor are not the fundamental degrees of freedom and that its fundamental equations must be derivable starting from a new paradigm based on the connection between the equations controlling the dynamics of the metric and the thermodynamics of horizons. This paradigm should make it possible to obtain the dynamical equations without being necessary to vary the metric in the principle of action;
6. The theory of Einstein is only an effective theory at low energy; thermodynamic description should provide keys to evaluate the corrections to this theory.

In the final chapter entitled *Gravity as an emergent phenomenon* of the book[28] which he has just published, Padmanabhan has reviewed the results that he obtained in the achievement of his program, in particular with regard to item #4: he has shown that it is possible to derive the equations of the gravitational field while varying, in a variational principle, some degrees of freedom other than the components of the metric and residing on the horizon, in agreement with the holographic principle which stipulates that "the true degrees of freedom of gravity for a volume $\mathcal{V}$, which cannot be eliminated by a gauge choice [i.e. by a choice of reference frame] reside on its border[29] $\partial\mathcal{V}$" He then shows that *in the volume delimited by the horizon, the cosmological constant is decoupled from gravity*. This decoupling is the consequence of to the fact that the equations of the field are invariant by change of the additive arbitrary constant of material energy, which gives the freedom to introduce the cosmological constant as a constant of integration once that the equations are solved and not in the action from which the equations derive. Such a scheme would make it possible to solve the problem of the cosmological constant and to lead to a satisfactory agreement with the observational data as they are consigned in the standard cosmological model.

---

[24] A. D. Sakharov, Sov. Phys. Dokl. 12, 1040 (1968)
[25] T. Jacobson, Phys. Rev. Letters 75, 1260 (1995)
[26] Pour une revue détaillée et de nombreuses références voir T. Padmanabhan, Thermodynamical Aspects of Gravity : New Insights, Rep. Prog. Phys. 73 046901 (2010) (arXiv:0911.5004)
[27] T. Padmanabhan, *Gravity as an emergent phenomenon : A conceptual description,* arXiv:0706.1654
[28] T. Padmanabhan, *Gravitation−Foundations and Frontiers−,* Cambridge University Press, The Edinburgh Building, Cambridge CB2 8RU, UK, (2010)
[29] T. Padmanabhan, *Gravity, a New Holographic Perspective,* arXiv :gr-qc/0606061

**5.2 Gravity as an entropic force**

Very recently, the gravity-thermodynamics connection caused a significant renewed interest following an article of Verlinde[30] in which he interprets gravity as an entropic force. He uses a heuristic reasoning, based on an analogy with the physics of polymers. He considers a polymer molecule immersed in a thermal bath, with one of its ends fixed in the bath. If one tries to extract the molecule by drawing it by the other end, it will be submitted to a force of entropic nature, which will tend to bring back the molecule in a state maximizing its entropy. Extending his analogical reasoning to the thermodynamics of the horizons, with a temperature and an entropy defined à la Hawking and Bekenstein, he succeeds in interpreting the force of gravity of Newton as an entropic force. He continues his reasoning by showing that the laws of Newton, that on inertia and that on the force of gravitation can be regarded as emergent, and that, in the same way, the principle of equivalence is emergent. Within a relativistic framework, he thus shows that his conjecture makes it possible to derive the Einstein's equations! The strong conclusion that he draws from his work is that it will be necessary to get accustomed with the idea that gravity is not a fundamental force: "It is time we not only notice the analogy, and talk about similarity, but finally do away with gravity as a fundamental force" He also suggests that such a paradigm shift should also take place in the string theory that is regarded as the best candidate for a quantum theory of gravitation, because the relation between open strings (which describe matter) and closed strings (which describe gravitation) can be also interpreted in terms of the emergence of gravitation.

## 6/ Conclusion

It thus appears to us today that current physics is in a prerevolutionary situation which is not without analogy with the one that prevailed at the beginning of the 20$^{th}$ century. New elements of observations like the rediscovery of the cosmological constant, the discovery of the acceleration of the cosmic expansion, the confirmation of the inflation scenario, as well as the interrogations caused by the dark matter, come to shake the contemporary theoretical building of physics. Just like what occurred at the beginning of the last century, statistical thermodynamics seems to be the missing piece of the puzzle. It makes it possible to bring closer this time two fields hitherto considered as disconnected: general relativity and quantum physics, each one resulting from the dissipation of one of the two clouds of Lord Kelvin. This time these two theories must cohabit in the description of black holes in particular, and cosmology in general. It is the holographic principle which casts the bridge between the principle of equivalence of general relativity and the principle of complementarity of quantum physics. It makes possible a description of the dynamics of black holes free from any paradox and provides the basis of an understanding of the cosmological term in the equation of Einstein. The holographic principle accounts for the echo on the horizon of the events occurring in the volume of the expanding universe. For any event horizon, this principle also applies.

The consequences of the existence of the holographic principle are of paramount importance because the central role of the horizon and of its thermal properties put into question the geometrical paradigm on which physics was built following the Greeks and since the 17$^{th}$ century. Information becomes first and space is emergent; the fine marble of the left hand side of the equation of Einstein is not more solid than the cheap and ordinary wood of its right hand side. The left hand side is to some extent the geometrical screen of the cave of Plato on which the shades of a more essential world are projected. The form which it takes in the equation of Einstein constitutes a low energy approximation of a more general theory yet to be elaborated.

This informational rather than geometrical approach at the same time makes it possible to comprehend the principle of Copernicus and the principle of relativity of Galileo. If space does not exist by itself, then there is no privileged position in space and any observer can be regarded as the center of the world observable by him (or her), or in other words, in absence of a viewpoint, the point

---

[30] E. Verlinde, *On the Origine of Gravity and the Laws of Newton*, (arXiv:1001.0785)

of view of an observer makes him (or her) the center of the world! The principle of inertia stating, according to the words of Galileo, that the inertial movement appears as "null" can now be re-interpreted in terms of information: as explained just above, binding acceleration to a variation of information exchanged by means of a non gravitational interaction (that is fundamental) amounts, according to the principle of equivalence, to binding gravity (that is emergent) to entropy.

With regard to the expansion of the universe, we would have entered, according to the current standard cosmological model, in a phase of re-inflation, controlled by the cosmological constant and leading, in the far future, to the hopelessly empty and uninteresting universe that we have hinted at the end of § 3.3 (see the figure and its legend). This aspect can appear discouraging. But we can somehow comfort ourselves because, as Pascal says it: "By space, the universe includes me and absorbs me like a point. By thought, I understand it[31]".

---

[31] « Par l'espace, l'univers me comprend et m'engloutit en un point. Par la pensée, je le comprends. » In french, the verb « comprendre » means to embed as well as to understand

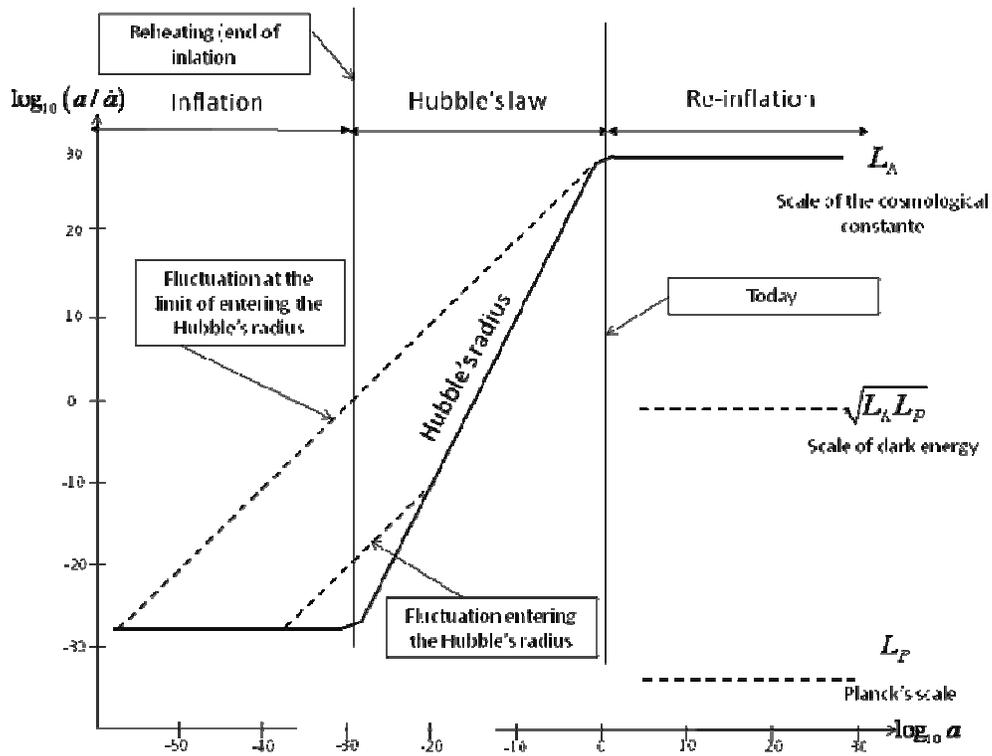

**The radius of the horizon as a function of the scale factor of the universe**

**Caption** This figure inspired by Padmanabhan (Astro-ph/0610492) shows in logarithmic scale and in natural units ($\hbar = c = 1$) the radius of the cosmological horizon ($a/\dot{a}$) as a function of the scale factor $a$. The phases of inflation (remote past) and re-inflation (remote future) correspond to de Sitter universes (implying horizons of constant radii), essentially void of matter and satisfying the perfect cosmological principle (namely universes everywhere and always identical to themselves). The phase obeying the Hubble's law $\left(a(t) \approx \sqrt{t} \Leftrightarrow a/\dot{a} \approx a^2\right)$ is the one described by the cosmology of concordance which describes "a brief history of matter"